\documentclass[a4paper,twocolumn]{article}

\usepackage[margin=2cm]{geometry}
\usepackage{graphicx}
\usepackage{caption}
\usepackage{authblk}
\usepackage[hidelinks]{hyperref}
\usepackage{xcolor}
\hypersetup{colorlinks, linkcolor={blue!50!black}, citecolor={blue!50!black}, urlcolor={green!25!black}}
\usepackage[super,numbers,sort&compress]{natbib}
\newcommand{\orgname}[1]{#1}
\newcommand{\state}[1]{#1}
\newcommand{\country}[1]{#1}
\newcommand{\email}[1]{\href{mailto:#1}{#1}}
\newcommand{\orgaddress}[1]{#1}
\newcommand{\orgdiv}[1]{#1}

\newcommand{\authormark}[1]{}

\newcommand\blfootnote[1]{%
	\begingroup
	\renewcommand\thefootnote{}\footnote{#1}%
	\addtocounter{footnote}{-1}%
	\endgroup
}

\newcommand{\thetitle}{A planar Airy beam light-sheet for two-photon microscopy}

\usepackage{todonotes}
\usepackage{upgreek}
\usepackage{wasysym}

\usepackage{hyperref}
\usepackage{xcolor}
\hypersetup{colorlinks, linkcolor={blue!50!black}, citecolor={blue!50!black}, urlcolor={green!25!black}}

\newcommand{\mm}{\;\mathrm{mm}}
\newcommand{\mum}{\;\upmu\mathrm{m}}
\newcommand{\nm}{\;\mathrm{nm}}

\newcommand{\muMol}{\;\upmu\mathrm{M}}
\newcommand{\mMol}{\;\mathrm{mM}}

\graphicspath{{figures/}}

\begin{document}

  \title{\thetitle}
  
  \newcommand{\firstauthor}{Neveen A. Hosny}
  \author[1]{\firstauthor}
  \author[1]{James A. Seyforth}
  \author[1]{Gunnar Spickermann}
  \author[1]{Thomas J. Mitchell}
  \author[1]{Pedro Almada}
  \author[2]{Robert Chesters}
  \author[2,3]{Scott J. Mitchell}
  \author[2]{George Chennell}
  \author[2,4]{Anthony C. Vernon}
  \author[2,3]{Kwangwook Cho}
  \author[2,4]{Deepak P. Srivastava}  
  \author[1]{Robert Forster}
  \author[5,6]{Tom Vettenburg*}
      
  \authormark{\firstauthor \textsc{et al}}
  
  \affil[1]{\orgdiv{M Squared Life}, \orgname{The Surrey Technology Centre}, \orgaddress{\state{Guildford, Surrey, GU2 7YG}, \country{UK}}}
  \affil[2]{\orgdiv{Basic and Clinical Neuroscience Dept}, \orgname{King's College London}, \orgaddress{\state{SE5 9NU}, \country{UK}}}
  \affil[3]{\orgdiv{UK Dementia Research Institute}, \orgname{King's College London}, \orgaddress{\country{UK}}}  
  \affil[4]{\orgdiv{MRC centre For Neurodevelopmental Disorders}, \orgname{King's College London}, \orgaddress{\country{UK}}}
  \affil[5]{\orgdiv{School of Physics and Astronomy}, \orgname{University of Exeter}, \orgaddress{\state{EX4 4QL}, \country{UK}}} 
  \affil[6]{\orgdiv{School of Science and Engineering}, \orgname{University of Dundee}, \orgaddress{\state{DD1 4HN}, \country{UK}}}
  \affil[*]{\email{t.vettenburg@dundee.ac.uk}}
  
  \maketitle
  
  \begin{abstract}
    We demonstrate the first planar Airy light-sheet microscope. Fluorescence light-sheet microscopy has become the method of choice to study large biological samples with cellular or sub-cellular resolution. The propagation-invariant Airy beam enables a ten-fold increase in field-of-view with single-photon excitation; however, the characteristic asymmetry of the light-sheet limits its potential for multi-photon excitation. Here we show how a planar light-sheet can be formed from the curved propagation-invariant Airy beam. The resulting symmetric light sheet excites two-photon fluorescence uniformly across an extended field-of-view without the need for deconvolution. We demonstrate the method for rapid two-photon imaging of large volumes of neuronal tissue.
    \blfootnote{\textbf{Abbreviations:} 2PE, two-photon excitation; 3D, three-dimensional; FOV, field-of-view; FWHM, full-width at half-maximum; EGFP, enhanced green fluorescent protein; NA, numerical aperture; PDMS, Polydimethylsiloxane.}
  \end{abstract}
  
  \section{Introduction}
    Fluorescence light-sheet microscopy has found rapid adoption in developmental biology and the neurosciences~\cite{Hillman19}.
    By illuminating a single plane, orthogonal to the detection axis of the microscope, light-sheet microscopy combines high resolution with unparalleled contrast and minimal sample exposure, thus limiting the potential for photo-bleaching and photo-toxic effects~\cite{Huisken04,Ahrens13}.
    However, diffraction limits the field-of-view~(FOV) in which the light-sheet illumination can be confined to a sufficiently thin section. This leads to a loss in contrast and axial resolution in all but the center of the FOV, while ineffectively exciting fluorescence elsewhere.
    Tiling multiple acquisitions~\cite{Fu16}, or swiping the focus of the light-sheet across the FOV may offer reprieve~\cite{Dean15,Chakraborty19}. Alternatively, propagation-invariant Bessel and Airy beams can be used to uniformly illuminate the complete FOV~\cite{Fahrbach10,Planchon11,Vettenburg14}. Light-sheets formed by such beams do have a transversal structure of side-lobes that would lead to poor axial resolution, unless the fluorescence excited by the side lobes is blocked~\cite{Fahrbach10}, or the main lobe is singled out using structured illumination or two-photon Bessel beam excitation~\cite{Planchon11}.
    
    \begin{figure*}[t]
    	\centerline{\includegraphics[width=\textwidth]{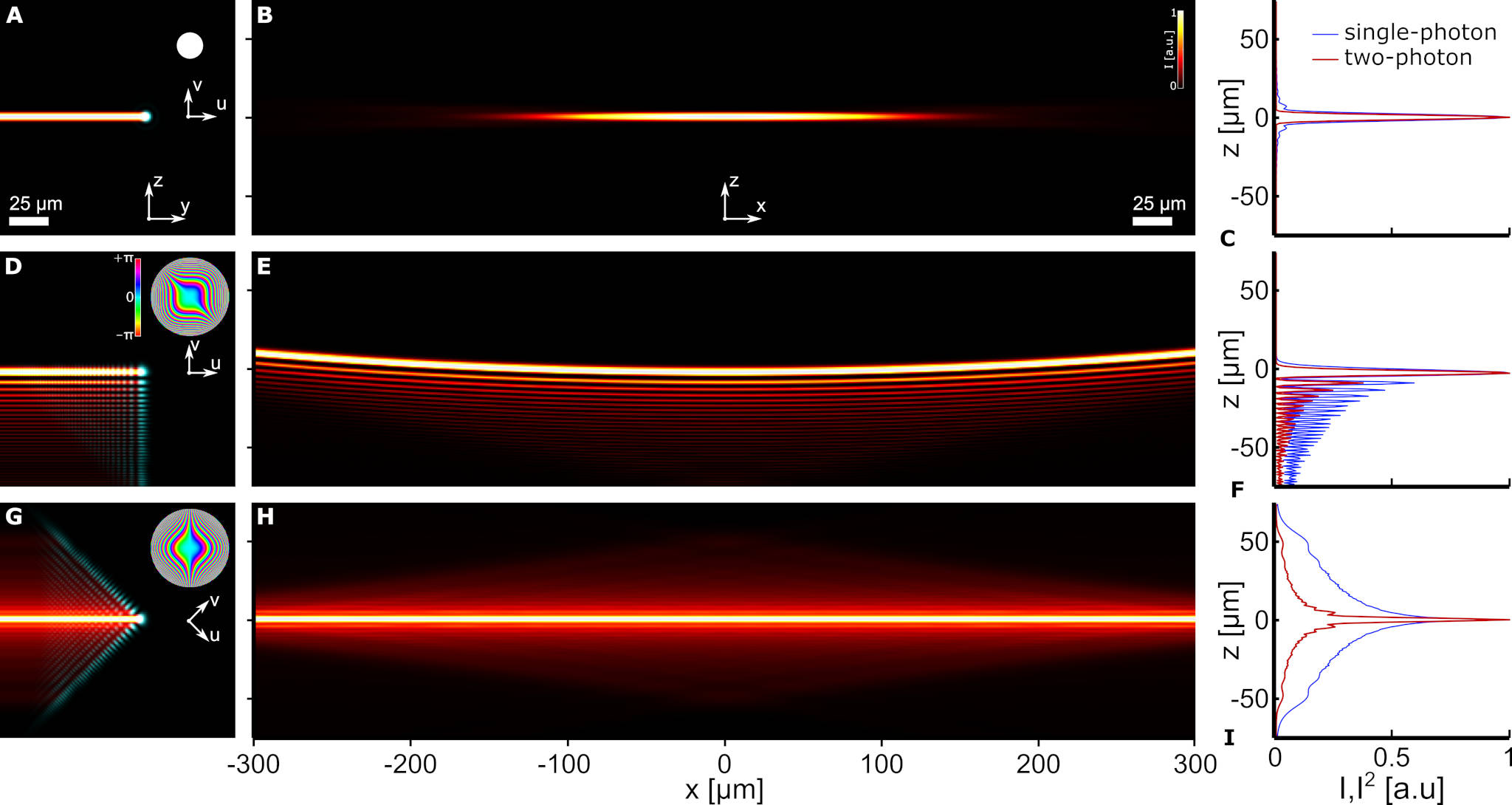}}
    	\caption{\label{fig:introduction}Depiction of 2PE light-sheet formation by rapidly scanning a truncated Gaussian beam (\textbf{A}-\textbf{C}), an Airy beam with aligned principal axes (\textbf{D}-\textbf{F}), and an Airy beam with its principal axes at $45^\circ$ to the light-sheet (\textbf{G}-\textbf{I}). The detection objective (not shown) is focused to the $x$-$y$-plane and two-photon excited fluorescence is collected along the $z$-axis. Transversal ($y$-$z$) cross-sections are shown in (\textbf{A},\textbf{D},\textbf{G}) for the beam waist (magenta) and the 2PE (false color legend in panel (\textbf{B})). Normalized linear intensity and 2PE for each light-sheet at its waist ($x=0$) are shown in panels (\textbf{C},\textbf{F},\textbf{I}). The inset disks show the relative back aperture size and phase (hue). The numerical apertures (NA) for the Airy beams is $0.30$, while that for the truncated Gaussian is chosen to be $0.10$ as a larger NA would have resulted in an impractically narrow FOV. The extents of the two-photon light-sheets can be seen in panels (\textbf{B},\textbf{E},\textbf{H}), which show false-color $x$-$z$-sections. (\textbf{B}) It can be seen that even at the lower NA, the truncated Gaussian only illuminates a fraction of the full FOV. (\textbf{E}) The Airy beam light-sheet illuminates the full FOV; however, the side-lobes of the Airy beam can be seen to form parallel, curved, surfaces. As a result, towards the extremes of the FOV, the main lobe does not coincide with the focal plane ($z=0$). (\textbf{H}) The $45^\circ$-rotated Airy beam can be seen to produce a single, planar, excitation surface. (\textbf{I}) By rotating the scanned beam with respect to the light-sheet plane, its side-lobes merge into a single structure (blue), which results in a highly-confined 2PE (red). Most importantly, the excitation coincides with the focal plane throughout the $0.60\mm$-wide FOV.}
    \end{figure*}
        
    Single-photon Airy light-sheet illumination can extend the FOV by an order of magnitude without requiring any physical filtering of the emitted fluorescence~\cite{Vettenburg14,Yang14}. The peculiar asymmetric transversal structure of the illumination enables digital-deconvolution to reconstruct 3D images with sub-cellular resolution. The image resolution is not determined by the width of the single-photon light-sheet but by the fine-structure of the illumination\cite{Vettenburg14}.
    
    Two-photon excitation can double the imaging depth~\cite{Truong11,Olarte12,Lavagnino13,Zong15}; furthermore, the longer wavelengths are less likely to interfere with the photo-receptor cells of the specimen~\cite{Wolf15}. While it has been shown that a two-photon Airy beam light-sheet can extend the FOV by a factor of six~\cite{Piksarv17}, the side-lobes in its transversal structure are suppressed. Although the width of the light-sheet is reduced, the loss in high spatial-frequency components in the two-photon excitation (2PE) profile precludes the digital recovery of the same high axial resolution as seen in its single-photon variant~\cite{Vettenburg14}.
    
    Notwithstanding, digital deconvolution is essential to correct for the asymmetry of the Airy beam light-sheet and the parabolic trajectory of the Airy beam that forms it~\cite{Siviloglou07a,Siviloglou07,Jeffrey08}. Without digital-deconvolution, the asymmetric transversal profile causes imaging artefacts, while the curvature of the light-sheet distorts the three-dimensional image formed by it. Here we show how the Airy beam light-sheet microscope can be modified to produce a uniform plane of illumination that obviates the need for deconvolution.
    
    \begin{figure*}[t]
    	\centerline{\includegraphics[width=\textwidth]{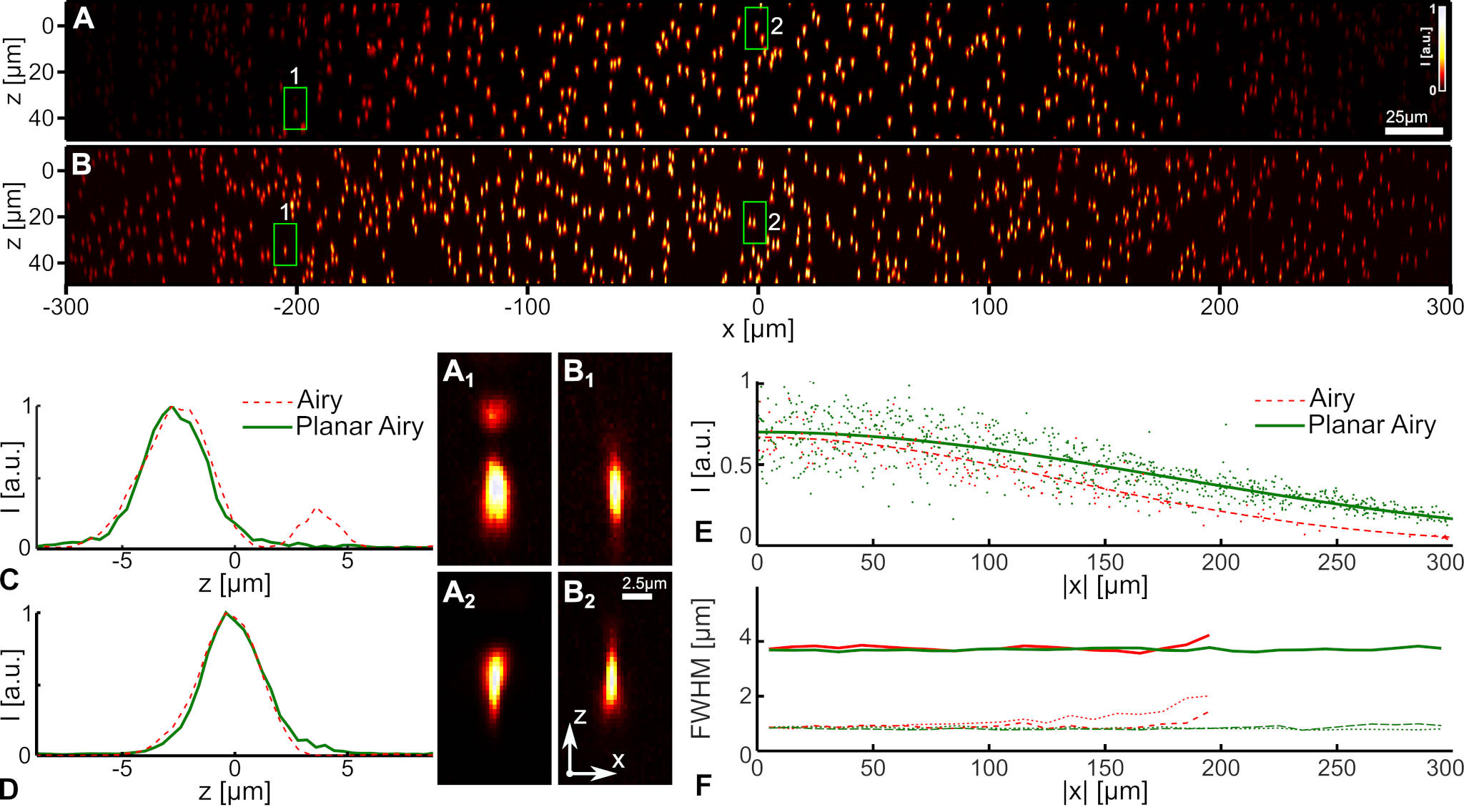}}
    	\caption{\label{fig:validation_and_analysis}Comparison of the two-photon Airy beam light-sheet and the two-photon Planar Airy light-sheet for uniformity and resolution. (\textbf{A},\textbf{B}) Maximum intensity projection of fluorescent microspheres with two-photon fluorescence excited using the conventional Airy light-sheet (\textbf{A}), and the planar Airy light-sheet (\textbf{B}). (\textbf{C},\textbf{D}) Intensity profiles of four images of isolated fluorescent microspheres near $x=-200\mum$ (\textbf{C}, \textbf{A$_1$}, \textbf{B$_1$}) and near the center of the FOV, $x=0\mum$ (\textbf{D}, \textbf{A$_2$}, \textbf{B$_2$}); for the conventional Airy light-sheet (red dashed) and the Planar Airy light-sheet (green solid). The locations of the sections are indicated with green rectangles in panels (\textbf{A}) and (\textbf{B}). All false-color images have been normalized to their maximum intensity for clarity. (\textbf{E}) Peak intensity as a function of distance to the FOV center $|x|$ for the microspheres illuminated with the conventional Airy light-sheet (red) and the Planar Airy light-sheet (green). To evaluate uniformity of the illumination, normal distributions have been fitted to the intensities and found to have full-width-at-half-maxima FOV of $311\pm17\mum$ (Airy, dashed red) and $415\pm11\mum$ (Planar Airy, green). (\textbf{F}) Full-width at half maximum in $x$ (dashed), $y$ (dotted), and $z$ (solid) of the microsphere images. Median values for every $10\mum$ interval are plotted for microspheres with peak intensities at least $50\%$ above the background.}
    \end{figure*}
    
  \section{The planar Airy light-sheet}
      A 2PE light-sheet can be generated by rapidly scanning a femto-second laser beam perpendicular to its propagation direction~\cite{Keller08a}. An extended `Airy light-sheet' is formed when, instead of a Gaussian, an Airy beam is used~\cite{Berry79,Vettenburg14,Piksarv17}. Although lasers can be designed to directly emit an Airy beam~\cite{Porat11}, any Gaussian beam can be converted into an Airy beam by introducing a cubic-polynomial phase modulation at a plane conjugate to the back aperture. This can be done using a diffractive spatial light modulator~\cite{Vettenburg14}, a phase mask, or off-the-shelf cylindrical lenses~\cite{Wang11b,Yang14}. The phase modulation introduces a position-dependent phase delay $\Delta\phi(u,v) \propto u^3+v^3$, where $u$ and $v$ are Cartesian coordinates normalized to the radius of the beam. The modulation depth is chosen to uniformly illuminate the FOV as discussed in Section~\ref{sec:methods}. For a single-photon Airy beam light-sheet it is advantageous to align the Cartesian axes ($u,v$) of the phase mask with those of the detection objective ($y,z$), as shown in the insets of Figures~\ref{fig:introduction}A~and~\ref{fig:introduction}D.
            
      Figure~\ref{fig:introduction}A shows the transversal profile of a 2PE truncated Gaussian beam (cyan) at $x=0$, as well as the light-sheet formed by scanning it along the $y$-axis (false-color, legend in panel Figure~\ref{fig:introduction}B). As can be seen from the $x$-$z$-cross-section of the light-sheet in Figure~\ref{fig:introduction}B, the truncated Gaussian light-sheet does not illuminate the full width of the FOV.
      Figure~\ref{fig:introduction}D shows the transversal profile of the 2PE Airy beam (cyan) at $x=0$ with its Cartesian axes, $u$ and $v$, parallel to the $y$ and $z$-axes. The inset in the top-right corner indicates the cubic phase modulation at the back aperture ($\mathrm{NA}=0.3$). Scanning the beam in the $y$-direction forms the Airy light-sheet (false-color). The $x$-$z$-section in Fig.~\ref{fig:introduction}E shows the characteristic parabolic trajectory of the Airy beam and the light-sheet it forms throughout the full FOV. Although the NA of the truncated Gaussian is only a third ($\mathrm{NA}=0.1$) of that of the Airy beam, it cannot match its uniformity. Multiple, high-contrast, side-lobes can be observed in the transversal structure parallel to the main intensity lobe of the Airy light-sheet. As can be seen from the comparison with the single-photon excitation (Fig.~\ref{fig:introduction}F), due to the quadratic scaling of the intensity, the transversal structure is less pronounced for 2PE~\cite{Vettenburg14}. Yet, the remaining side-lobes and the curvature of the light-sheet introduce imaging artefacts and distortion.    
      A straightforward modification avoids this, thus eliminating the need for deconvolution and image processing.
      
      Figures~\ref{fig:introduction}(G,H), show how a planar light-sheet can be formed by rotating the phase modulation $45^\circ$ around the propagation axis, $x$, so that the parabolic trajectory of the Airy beam coincides with the $x$-$y$-plane, the focal plane of the detection objective. Furthermore, such planar light-sheet can be seen to have a $z$-symmetry in its transversal structure, with side-lobes that overlap and blur into a triangular excitation profile (Fig.~\ref{fig:introduction}I). This negates the need for deconvolution of Airy side lobes and geometric correction of the curved Airy beam. Moreover, since the beam is no longer curved, the depth-of-field of the detection objective is not limited by the illumination, thus allowing high NA detection objectives for optimal lateral resolution.
      
  \section{Results and Discussion}
	
    A 2PE Airy light-sheet was produced by introducing a cubic phase mask in the illumination path of an inverted light-sheet microscope (iSPIM\cite{Wu11}). The cubic phase mask was mounted in a rotation mount to facilitate switching between the conventional ($0^\circ$) and the planar Airy light-sheet ($45^\circ$). While the former configuration has been preferred for single photon excitation and deconvolution~\cite{Vettenburg14}; here, we show that the planar Airy light-sheet is highly advantageous when using two-photon excitation.
    
    To quantify its imaging capabilities, a $0.60\mm$-wide volume of fluorescent microspheres ($\diameter 500\nm$) was imaged using the conventional 2PE Airy light-sheet (Fig.~\ref{fig:validation_and_analysis}A) and the planar Airy light-sheet (Fig.~\ref{fig:validation_and_analysis}B). Although the NA of the illumination was identically $0.30$ in both cases, it is apparent that the illumination of the Airy light-sheet is less uniform than that of the planar Airy light-sheet. To quantify the widths of the FOV, the peak intensity of the microspheres was plotted as a function of the absolute distance, $|x|$, from the center of the field-of-view (Fig.~\ref{fig:validation_and_analysis}E) for the Airy light-sheet (red) and the planar Airy light-sheet (green). To avoid counting overlapping point-spread functions, only clearly isolated (in 3D) microspheres were considered. A Gaussian curve was fitted to determine the full-width-at-half-maximum (FWHM) of the illuminated FOV. The 2PE Airy light-sheet illuminates a FOV with a FWHM of approximately $311\pm17\mum$, while the planar Airy light-sheet extends this by a third to $415\pm11\mum$. The improved uniformity of the illumination also suggests that the planar Airy light-sheet can enable a reduction in laser power and sample exposure.
    
    \begin{figure}[t]
    	\centerline{\includegraphics[width=\columnwidth]{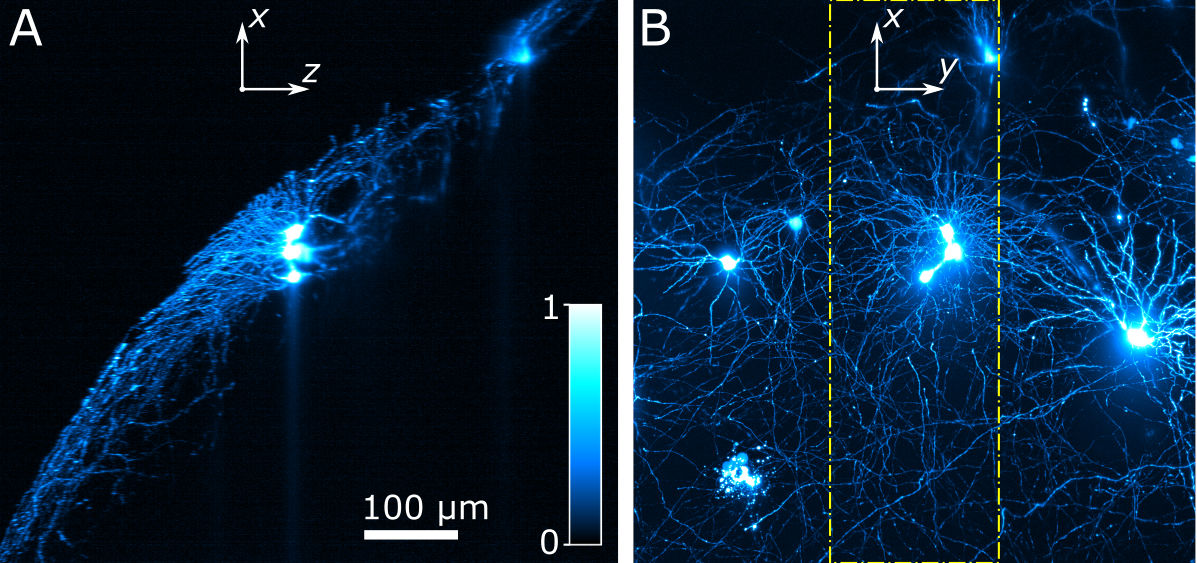}}
    	\caption{\label{fig:rat_neurons}Maximum intensity projection of $0.60 \times 0.60 \times 0.60\mm^3$ of Wistar rat hippocampal tissue, imaged using the two-photon planar Airy light-sheet. (\textbf{A}) The $x$-$z$-projection along the $y$-axis of the volume indicated by the yellow dash-dotted box in panel (B). The variation in axial ($z$) resolution can be seen to be minimal. (\textbf{B}) The $x$-$y$-projection along the $z$-axis, showing high lateral resolution throughout the field-of-view.}
    \end{figure}

    A closer examination of the images of the individual microspheres provides insight into the lack of uniformity of the 2PE Airy light-sheet illumination. Figures~\ref{fig:validation_and_analysis}A$_1$~and~\ref{fig:validation_and_analysis}B$_1$ show microspheres near $x=-200\mum$ for the conventional 2PE Airy light-sheet and the planar Airy-light microscope, respectively. The exact location of the sub-volumes is indicated with with a green rectangle in Figures~\ref{fig:validation_and_analysis}A~and~\ref{fig:validation_and_analysis}B. Two lobes of the transversal structure are distinctly visible for the conventional 2PE Airy light-sheet (Fig.~\ref{fig:validation_and_analysis}A$_1$); while the planar Airy light-sheet produces a single, compact, point-spread function (Fig.~\ref{fig:validation_and_analysis}B$_1$). Cross-sections of the microspheres are shown in Figure~\ref{fig:validation_and_analysis}C for clarity. At $|x|\approx 200\mum$, both the main and the second intensity lobe of the transversal profile are within the depth-of-field of the detection objective for the conventional Airy light-sheet, though neither is in optimal focus. In contrast, the planar Airy light-sheet only has a single well-defined plane of high intensity that coincides with the focal plane of the detection objective (Fig.~\ref{fig:introduction}I). The images of microspheres are therefore relatively independent of the their position in the FOV, as can be seen by comparing Figures~\ref{fig:validation_and_analysis}B$_1$~and~\ref{fig:validation_and_analysis}B$_2$.
    
    As an indication of resolution, the FWHM in the three dimensions and its median plotted for every $10\mum$-interval in Figure~\ref{fig:validation_and_analysis}F. The median FWHM in $(x,y,z)$ over the FOV are found to be $(0.91, 1.02, 3.74)\mum$ for the conventional Airy light-sheet and $(0.81, 0.85, 3.69)\mum$ for the Planar Airy light-sheet, respectively. 
    Note that insufficient bright isolated references were found beyond $|x|\geq200\mum$ for the conventional Airy light-sheet, while the planar Airy light-sheet ensures that all microscopheres near the focal plane are well-illuminated.
    Although these values are an upper bound due to the finite diameter of the microspheres, it is clear that the resolution is relatively constant throughout the FOV for the planar Airy light-sheet (Fig.~\ref{fig:validation_and_analysis}F, green).
    
    %
    %
    
    The lack of side-lobes and curvature of the planar Airy light-sheet simplifies the imaging process and removes several constraints. No deconvolution, nor geometric correction is required with the planar Airy light-sheet. In turn, this obviates Nyquist sampling in the axial dimension ($\Delta z=0.4\mum$), thereby enabling faster volumetric recording. The numerical aperture of the detection objective can be chosen so that its Rayleigh range matches the planar Airy light-sheet's transverse profile. 
    
    Fig.~\ref{fig:rat_neurons} shows Venus-fluorescence expressing neurons from organotypic cultured hippocampal slices of male Wistar rat. Two-photon Planar Airy light-sheet imaging enables the rapid visualization of synaptic function and microstructure with high resolution in live tissue, thus facilitating investigations into the relationship between neuronal structure and function.
    The neurons extend over the $x \times y \times z = 0.60\times 0.60\times 0.66\mm^3$ imaging volume. Panels (A) and (B) show $y$ and $x$-axis projections, respectively. It can be noted that the two-photon excitation does not vary appreciably from the center to the edges of the $0.60\mm$-wide FOV.
    
    \begin{figure}[t]
    	\centerline{\includegraphics[width=\columnwidth]{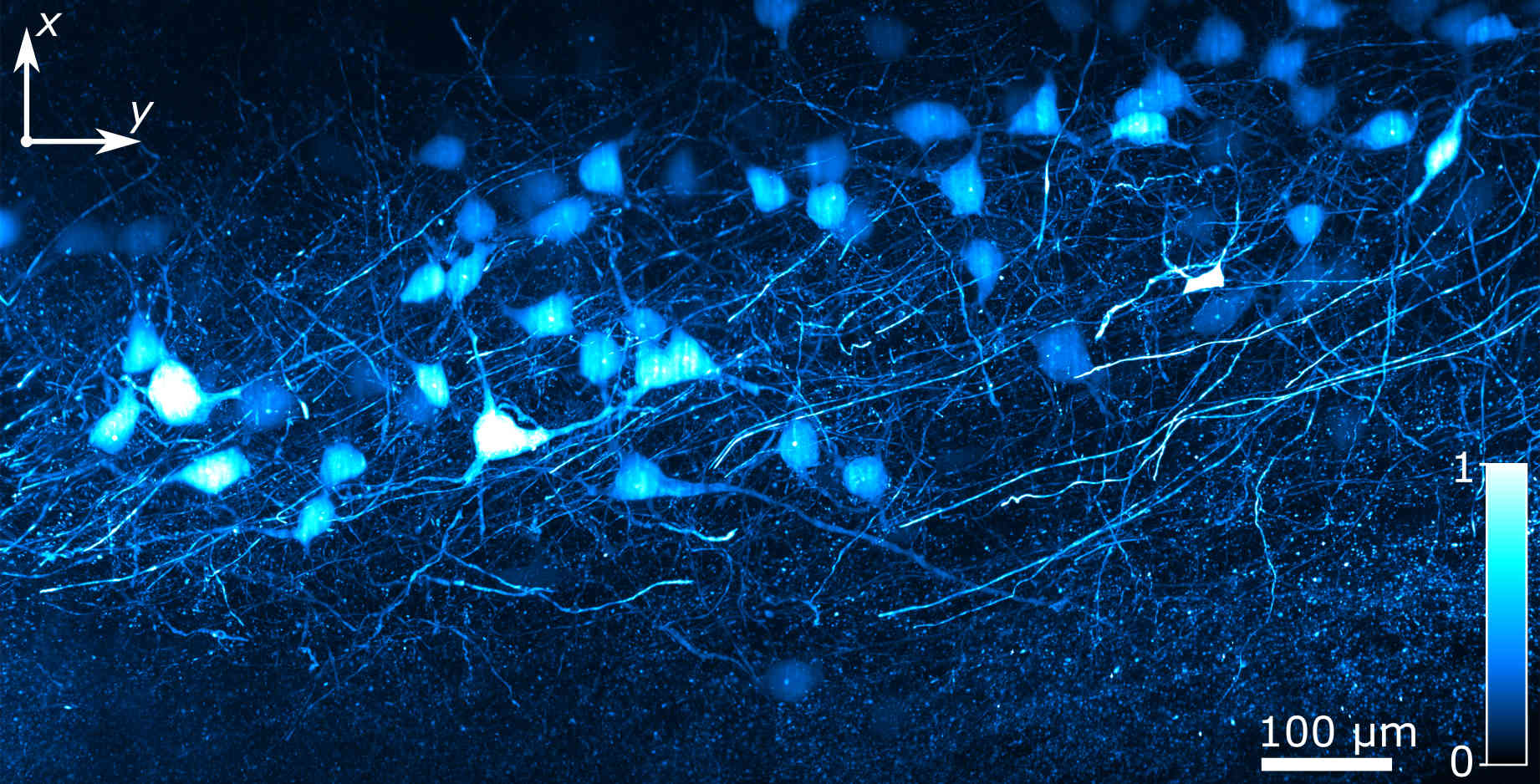}}
    	\caption{\label{fig:mouse_neurons}Maximum intensity projection of $x \times y \times z = 0.60 \times 1.15 \times 0.60\mm^3$ of mouse brain tissue imaged using the two-photon planar Airy light-sheet.} 
    \end{figure}

    Larger volumes can be imaged by increasing the scan distance along the $z$-axis or by tiling multiple acquisition volumes in the $x$ or $y$-direction. The latter is demonstrated by acquiring neighboring $0.60\times 0.60\times 0.60\mm^3$ volumes of cleared mouse brain tissue. Two side-by-side stacks were acquired to produce a $y = 1.15\mm$-tall 3D data cube. Maximum intensity projections are shown in Fig.~\ref{fig:mouse_neurons}. Each stack consists of $500$ slices, with a slice-spacing of $1.2\mum$, and illuminated with the planar Airy light-sheet for a duration of $50\;\mathrm{ms}$. This resulted in an acquisition time of $25\;\mathrm{s/stack}$. 
      
  \section{Conclusion}
    We have demonstrated how a planar Airy light-sheet can be realized based on the propagation-invariant, yet curved, Airy beam. The symmetric intensity profile of the planar Airy light-sheet eliminates any requirement for deconvolution and image processing. We characterized the performance of the planar Airy light-sheet microscope and demonstrate how it can be effective for rapid imaging of large volumes of neuronal tissue with two-photon excitation. 
    The propagation-invariant Airy beam can potentially be generated by combining low-cost off-the-shelf cylindrical lenses~\cite{Papazoglou10}.
    We anticipate that the advantages of this method will be further enhanced with higher-order non-linearities to image deeper into tissue~\cite{EscobetMontalban18}.
  
  \section{Methods}\label{sec:methods}
    \subsection{Optical set-up and image formation}
      An Airy beam light-sheet microscope in iSPIM configuration~\cite{Keller08a,Wu11,Vettenburg14} is modified to enable the axial rotation of the beam shaping optics. A cubic phase modulation introduces a phase delay, $\Delta\phi(u,v) = 2\pi\alpha \left(u^3+v^3\right)$, in the illumination beam before reimaging it to the back aperture of the illumination objective (Olympus $10\times$ NA 0.30). Here, $u$ and $v$ are Cartesian coordinates, centered at the back-aperture and normalized to its radius, while $\alpha$ is a unit-less constant that is approximately proportional to the propagation invariance of the light-sheet~\cite{Vettenburg14}. Here, the value of $\alpha$ was determined to be $10.2$ at a wavelength of $930.9\nm$ for 2PE. 
      To enable switching between a conventional Airy light-sheet and a planar Airy light-sheet, the phase modulation can be rotated by $45^\circ$ around the optical axis.
      Fluorescence emission is collected using a second water dipping objective (Olympus $20\times$ NA 0.50), orthogonal to the excitation plane, and refocused onto the sensor array (Orca Flash 4.0 v2). Using a galvanometer mirror, the Airy beam is scanned along the $y$-axis during the acquisition of each single-plane image. A three-dimensional volume is acquired by motorized translation of the sample orthogonally to the focal plane.
      
      Two-photon excitation was achieved using a mode-locked Ti:Sapphire mode-locked laser (Sprite XT, M Squared Lasers, UK), at a wavelength of $\lambda=930.9\nm$, a pulse duration of $140\;\mathrm{fs}$ and a repetition rate of $80\;\mathrm{MHz}$.
      
      No geometric correction or image deconvolution is used. Raw data is analyzed and visualized using Matlab (MathWorks, USA) and FIJI~\cite{Schindelin12}. Imaris Image Analysis Software (Oxford Instruments, UK) to create a three-dimensional visualization.
      
    \subsection{Sample Preparation}
      \subsubsection{Fluorescent microspheres}
      	To experimentally verify the resolution and uniformity, image stacks of a phantom sample were acquired. Fluorescent microspheres ($\diameter 500\nm$ Tetraspeck, Thermofisher UK) were sparsely suspended in low melting point $1.2\%$ agarose (Ultrapure, Invitrogen), loaded in to a sample chamber and immersed in water. Using the beads allowed the resolution and brightness to be evaluated as a function of the position in the FOV.
      
      \subsubsection{Biological samples}
      	All procedures were carried out in accordance with the UK Animals (Scientific Procedures) Act, 1986. All animal experiments were given ethical approval by the ethics committee of King's College London (UK).
      	
      	The biological samples used in the experiments were held in place with a thin layer of low melting point $1.2\%$ agarose on Polydimethylsiloxane (PDMS) plinths inside the samples chambers.
      	
      	\textbf{Mouse brain tissue.}
      	Brain tissue from adult male mice expressing Thy-1-GFP (Tg(Thy1-EGFP)MJrs/J; a generous gift from Professor Robert Hindges, King's College London) was fixed and rendered optically transparent using passive CLARITY~\cite{Tomer14}.
      	
      	\textbf{Organotypic hippocampal slice culture and transfection.}
      	Male 7-day old Wistar rats (Charles River, UK) were used to prepare organotypic hippocampal slices for live imaging.
      	The organotypic hippocampal slice culture (Fig.~\ref{fig:rat_neurons}) was prepared as previously described~\cite{Yi18}. All steps were carried out under sterile conditions. Briefly, schedule 1 procedure was performed, rat brains rapidly removed and placed into ice-cold dissecting medium containing: $238\mMol$ sucrose, $2.5\mMol$ KCl, $26\mMol$ NaHCO3, $1\mMol$ NaH2PO4, $5\mMol$ MgCl2, $11\mMol$ D-glucose and $1\mMol$ CaCl2. Hippocampi were removed and transverse hippocampal slices ($350\muMol$) were cut. Following washing, slices were placed upon sterile, semi-porous membranes (Millipore, USA) and stored at the interface between air and culture medium containing: $78.8\%$ minimum essential medium with L-glutamine, $20\%$ heat-inactivated horse serum, $30\mMol$ HEPES, $26\mMol$ D-glucose, $5.8\mMol$ NaHCO3, $2\mMol$ CaCl2, $2\mMol$ MgSO4, $70\muMol$ ascorbic acid and $1\;\upmu\mathrm{g}\;\mathrm{mL}^{-1}$ insulin (pH adjusted to $7.3$ and 320 - 330 mOsm $\mathrm{kg}^{-1}$). The slices were then cultured in an incubator ($35\;^\circ\mathrm{C}$, $5\%$ CO2) for $7-10$ days in vitro (DIV). The medium was changed every $2$ days. Neurons were transfected with mVenus using a biolistic gene gun (Helios Gene-gun system, Bio Rad, U.S.A.) at DIV 4. Imaging assays were performed $5$ days after transfection (DAT)~\cite{Yi17}. The mVenus fluorescent protein was expressed throughout the cells to allow visualization of the neuronal architecture.
      
  \section*{Acknowledgments}
    We would like to thank Professor Robert Hindges, King's College London, for the generous gift of the Thy1-GFP expressing mice; as well as the Wohl Cellular Imaging Centre for their help with this study.
    
    This project is supported by the Open Innovation Funding Platform award (University of Exeter, UK) for the project "Novel illumination strategies for improved light sheet microscopy" (PS-CEMPS-4786). TV is a UKRI Future Leaders Fellow supported by grant MR/S034900/1.
    
  \subsection*{Author contributions} 
    NAH and JAS designed and conducted the experiments, GS constructed the optical system and general control, TJM processed the rat hippocampal data, PA designed the control software for the multi-photon laser and lightsheet. RF supervised NAH, JAS, GS, TJM, and PA and coordinated. RC produced the mouse brain tissue under supervision of ACV and DPS, SJM prepared the hippocampal cell cultures for imaging under supervision of KC. Animal Tissue images were taken by GC. NAH, GS, TJM, and TV analyzed the data, created the figures, and wrote the paper. All authors reviewed the manuscript.

%
  
  \bibliography{../../docs/Common}
    
\end{document}